\documentclass[acmsmall]{acmart}

\acmConference[FSE 2024]{International Symposium on the Foundations of Software Engineering}{15 -- 19 July 2024}{Porto de Galinhas, Brazil}

\setcopyright{acmlicensed}
\acmDOI{10.1145/3643774}
\acmYear{2024}
\copyrightyear{2024}
\acmSubmissionID{fse24main-p1207-p}
\acmJournal{PACMSE}
\acmVolume{1}
\acmNumber{FSE}
\acmArticle{48}
\acmMonth{7}
\received{2023-09-29}
\received[accepted]{2024-01-23}

\usepackage{tcolorbox}
\usepackage[utf8]{inputenc}
\usepackage{array,multirow,etoolbox,graphicx,subcaption,fancyhdr,tabularx,longtable}
\usepackage{float}
\usepackage{graphicx}
\usepackage{amsmath} 
\usepackage{tcolorbox}
\usepackage{booktabs}
\usepackage{mathtools}
\usepackage{xcolor}
\usepackage{hyperref}
\newtcolorbox{TcolorBox}[1]{fonttitle=\bfseries,title=#1}
\usepackage{csquotes}
\usepackage[euler]{textgreek}
\usepackage{microtype}
\usepackage{tablefootnote}
\usepackage{threeparttable}
\usepackage{array,multirow,etoolbox,graphicx,subcaption,fancyhdr,tabularx,longtable}
\usepackage{makecell}
\usepackage{pgfplots}
\usepackage{tablefootnote}
\usepackage{changepage}

\usepackage{xurl} 
\usepackage{tabularx}
\usepackage{caption}
\usepackage{subcaption}
\usepackage{xspace}

\usepackage{enumitem}
\newlist{steps}{enumerate}{1}
\setlist[steps, 1]{label = Step \arabic*:}

\usepackage{xspace}
\usepackage{dirtytalk}
\usepackage{xcolor}

\newcommand{\cc}{\textsc{CodeCompose}\xspace}
\newcommand{\company}{Meta\xspace}

\usepackage{xspace}
\usepackage{dirtytalk}
\usepackage{xcolor}

\definecolor{teal}{RGB}{0,128,128}

\newcommand{\todo}[1]{\textcolor{red}{{\it [TODO: #1]}}}
\newcommand{\etal}{\emph{et al.}\xspace}

\newcommand{\DefMacro}[2]{\expandafter\newcommand\csname rmk-#1\endcsname{#2}}
\newcommand{\UseMacro}[1]{\csname rmk-#1\endcsname}
\newcommand{\rom}[1]{\uppercase\expandafter{\romannumeral #1\relax}}
\newcommand{\code}[1]{\texttt{#1}}

\newcommand{\LearningRate}{$5\text{e-}4$\xspace}
\newcommand{\BatchSizePerDevice}{20\xspace}
\newcommand{\BatchSize}{2.5 thousand\xspace}
\newcommand{\NumTrainingGPUs}{128\xspace}
\newcommand{\NumInferenceGPUs}{150\xspace}
\newcommand{\NumDaysToTrain}{4\xspace}
\newcommand{\GPUType}{A100\xspace}
\newcommand{\NumEpochs}{4\xspace}

\newcommand{\numLangs}{9\xspace}

\newcommand{\numDevsActivity}{16 thousand\xspace}
\newcommand{\numSuggestionsActivity}{4.5 million\xspace}
\newcommand{\accRateActivity}{22\%\xspace}
\newcommand{\percCodeActivity}{8\%\xspace}

\title{AI-assisted Code Authoring at Scale: Fine-tuning, deploying, and mixed methods evaluation}

\begin{document}

\begin{abstract}

Generative LLMs have been shown to effectively power AI-based code authoring tools that can suggest entire statements or blocks of code during code authoring.
In this paper we present \cc, an AI-assisted code authoring tool developed and deployed at \company internally.
\cc is based on the InCoder LLM that merges generative capabilities with bi-directionality.
We have scaled up \cc to serve tens of thousands of developers at \company, across 9 programming languages and several coding surfaces.
We present our experience in making design decisions about the model and system architecture for \cc that addresses these challenges.

To release a LLM model at this scale, we needed to first ensure that it is sufficiently accurate. In a random sample of 20K source code files, depending on the language, we are able to reproduce hidden lines between 40\% and 58\% of the time, an improvement of $1.4\times$ and $4.1\times$ over a model trained only on public data. 

We gradually rolled \cc out to developers. At the time of this writing, 16K developers have used it with 8\% of their code coming directly from \cc. 

To triangulate our numerical findings, we conduct a thematic analysis on the feedback from 70 developers. We find that 91.5\% of the feedback is positive, with the most common themes being discovering APIs, dealing with boilerplate code, and accelerating coding.
\company continues to integrate this feedback into \cc.

\end{abstract}

\ccsdesc[500]{Computing methodologies~Neural networks}
\ccsdesc[500]{Software and its engineering}

\keywords{AI, Developer productivity, Neural code completion, Program synthesis}

\author{Vijayaraghavan Murali}
\email{vijaymurali@meta.com}
\orcid{0009-0009-1374-7334}
\affiliation{
  \institution{Meta Platforms Inc.}
  \country{USA}
}
\author{Chandra Maddila}
\email{cmaddila@meta.com}
\orcid{0000-0002-9432-1045}
\affiliation{
  \institution{Meta Platforms Inc.}
  \country{USA}
}
\author{Imad Ahmad}
\email{imadahmad@meta.com}
\orcid{0009-0005-8606-3946}
\affiliation{
  \institution{Meta Platforms Inc.}
  \country{USA}
}
\author{Michael Bolin}
\email{mbolin@meta.com}
\orcid{0009-0002-6955-1970}
\affiliation{
  \institution{Meta Platforms Inc.}
  \country{USA}
}
\author{Daniel Cheng}
\email{danielcheng@meta.com}
\orcid{0009-0007-1708-8132}
\affiliation{
  \institution{Meta Platforms Inc.}
  \country{USA}
}
\author{Negar Ghorbani}
\email{negargh@meta.com}
\orcid{0000-0002-0528-6138}
\affiliation{
  \institution{Meta Platforms Inc.}
  \country{USA}
}
\author{Renuka Fernandez}
\orcid{0009-0007-4734-4328}
\email{rdfernandez@meta.com}
\affiliation{
  \institution{Meta Platforms Inc.}
  \country{UK}
}
\author{Nachiappan Nagappan}
\email{nnachi@meta.com}
\orcid{0000-0003-1358-4124}
\affiliation{
  \institution{Meta Platforms Inc.}
  \country{USA}
}

\author{Peter C. Rigby}
\orcid{0000-0003-1137-4297}
\affiliation{%
  \institution{Meta Platforms Inc.}
  \country{USA}
}
\affiliation{%
  \institution{Concordia University}
  \country{Canada}
}
\email{pcr@meta.com}

\renewcommand{\shortauthors}{V Murali, C Maddila, I Ahmad, M Bolin, D Cheng, N Ghorbani, R Fernandez, N Nagappan, PC Rigby}
\maketitle




\section{Introduction} \label{intro}

Recently, large language models (LLMs)\cite{vaswani2017attention,brown2020language,devlin2019bert} have exhibited the ability to assimilate vast amounts of knowledge after being trained on the data in various corpora.
This characteristic of LLMs has made them highly impactful in assisting developers in authoring code\cite{copilot,codewhisperer,googleblog,nijkamp2023codegen}.
Specifically, when they are trained to predict the next token in a code sequence, LLMs can become powerful coding assistants that can suggest entire statements or blocks of code during code authoring.
Such a coding assistant trained on an organization's code repository can surface internal knowledge during code authoring when developers are most likely to seek that information.

At \company, we built an AI-assisted code authoring system named \cc to explore the application of LLM technology for code authoring.
\company is a large software company with a complex code base that covers a wide range of applications from social networking and virtual reality to software engineering infrastructure, such as continuous integration tooling and workplace coordination. Tens of thousands of developers work on billions of lines of code (LOCs) in a monolithic repository that hosts source code written in 9+ programming languages. One of the prominent activities in the Software Development Life Cycle (SDLC) at \company is code authoring. A significant amount of contextual knowledge on the internal software development processes, and libraries, is embedded in the source code or is confined to a small set of developers. For example, a question like \emph{``How can I upload a table to Hive in Hack?''} can be answered using that knowledge.
The state of the source code and the developers associated with the source code keeps evolving constantly - internal libraries and tools get added and deprecated, while developers move across teams and change roles.
At \company's scale, keeping up with the knowledge required to accomplish coding tasks is challenging. Additionally, the dynamic environment at a large software company like \company poses interesting challenges concerning knowledge discovery and achieving developer efficiency. 

\begin{figure}
\begin{tabular}{c}
\includegraphics[width=0.8\columnwidth]{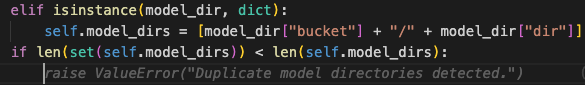} \\ (a) \\ \\
\includegraphics[width=0.8\columnwidth]{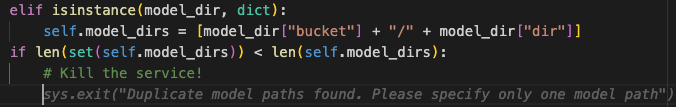} \\ (b) \\ \\
\includegraphics[width=0.5\columnwidth]{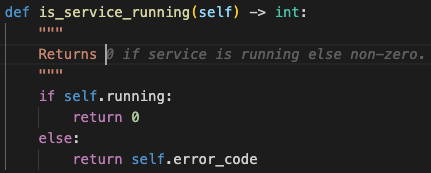} \\ (c) \\
\end{tabular}
\caption{\cc (a) offers inline code suggestions in VSCode in a grey text appearing after the cursor when the user is typing code (Tab to accept), (b) changes its suggestion to adapt to a natural language comment, (c) suggests code or documentation based on code below the current position.}
\label{fig:cc-screenshots}
\end{figure}

Over the past year, we (i) conducted R\&D on the LLM (model) architecture, training data, and training objective, (ii) built an end-to-end coding assistant that offers code suggestions in various code authoring surfaces across the company, and (iii) gathered quantitative metrics and qualitative feedback to measure the impact of the \cc system.
\cc has the following characteristics:
\begin{itemize}[leftmargin=*]
    \item Multi-lingual: \cc is based on the InCoder LLM~\cite{fried2023incoder} and has been trained on 9+ programming languages at \company. As such, it inherits the property of LLMs to be multi-lingual.
    
    \item Customized for the organization: \cc is fine-tuned on \company's internal code repository, and is thus able to handle languages such as Hack and Flow. \cc is also deeply integrated with \company's version of VSCode, as seen in Figure~\ref{fig:cc-screenshots}(a), and other editors.
    
    \item Natural language proficiency: \cc can understand inline comments in natural language and generate code that adheres to the comment, as shown in Figure~\ref{fig:cc-screenshots}(b). It can also fluently generate comments, messages, and documentation.
    
    \item Bi-directional: \cc can look beyond the cursor to suggest code at the current position. In Figure~\ref{fig:cc-screenshots}(c) it can be seen suggesting the docstring for a function from code that conventionally appears after the docstring.
\end{itemize}

In this paper we present how we built \cc, discuss the unique challenges of our scale and how they influenced design decisions about \cc, and finally, results from an extensive large-scale deployment including developer feedback on the impact of \cc on how code is authored at \company. 
Specifically, we provide evidence related to the following research questions:

\textbf{RQ1. Model Evaluation: How well does \cc generate one hidden line of code from existing code snippets?}

Before designing and releasing a system to engineers, we test it on the existing code base. 
We randomly mask part of the code 
and allow the model to predict lines in the masked part to evaluate its accuracy. The code before and some of the code after the masked part is also fed to the model as the context, along with some additional metadata. We contrast the public InCoder model~\cite{fried2023incoder}, against our fine-tuned model with both the Causal Masking (CM) and Language Causal Masking (LCM) objectives.

\textit{Result Summary:} 
We find that LCM \cc model exactly recreates the masked line between 40\% and 58\% of the time and has a high BLEU score between 56\% and 73\%, depending on the programming language. This represents an improvement over the Public model between $1.4\times$ and $4.1\times$.

\textbf{RQ2. Adoption: How many suggestions are accepted by engineers and what proportion of the code is written by \cc?}

Few studies of LLM for coding are able to go beyond backtests. There are some blog posts demonstrating the use by professional engineers. At \company we are able to gradually rollout new features to engineers. 
We progressively roll \cc out and monitor its impact through the number of accepted suggestions and the proportion of code that is written by \cc.

\textit{Result Summary:} In this paper, we make over 4.5M suggestions to 16K engineers. We see an acceptance rate of 22\% which is comparable to those at Google and GitHub. We see that 8\% of the changed code is written by \cc, which is higher than the 3\% reported by Google. \cc is rolled out to 100\% of engineers at \company.

\textbf{RQ3. Developer Feedback: How do developers perceive \cc in their daily work?}

We triangulate our numerical analysis, which show evidence of adoption, with feedback from developers and discussion of actual usages~\cite{Creswell2017MixedMethods}.
Using the feedback group for \cc, we conduct a thematic analysis of the comments from 70 engineers. We provide direct quotes from engineers to contextualize and ground the themes we discovered.  We also use these themes and feedback to iteratively improve \cc.

\textit{Result Summary:} 
The feedback was overwhelmingly positive, with only 8.5\% of comments containing negative feedback and less than 1\% of engineers turning off \cc. The top two feedback comment themes related to discovering APIs and dealing with boilerplate (23\%), and accelerating coding (20\%). \company continues to invest in \cc efforts to iteratively improve the AI-assisted authoring experience.

The paper is organized as follows. Section~\ref{model} presents details about the \cc model. Section~\ref{design} presents the system architecture. Section~\ref{eval} presents results from our large-scale deployment at \company. Section~\ref{threats} discusses threats to validity. Section~\ref{related} discusses related work. Section~\ref{sec:challenges} discusses the challenges for coding assistants at scale. Section~\ref{conclusion} concludes the paper.

\section{Model Development and Evaluation Methodology}
\label{model}
\label{sec:methodModel}

In this section, we provide details about the underlying LLM architecture for \cc, and discuss considerations that affected our decision.

\begin{figure*}
    \centering
    \includegraphics[width=\textwidth]{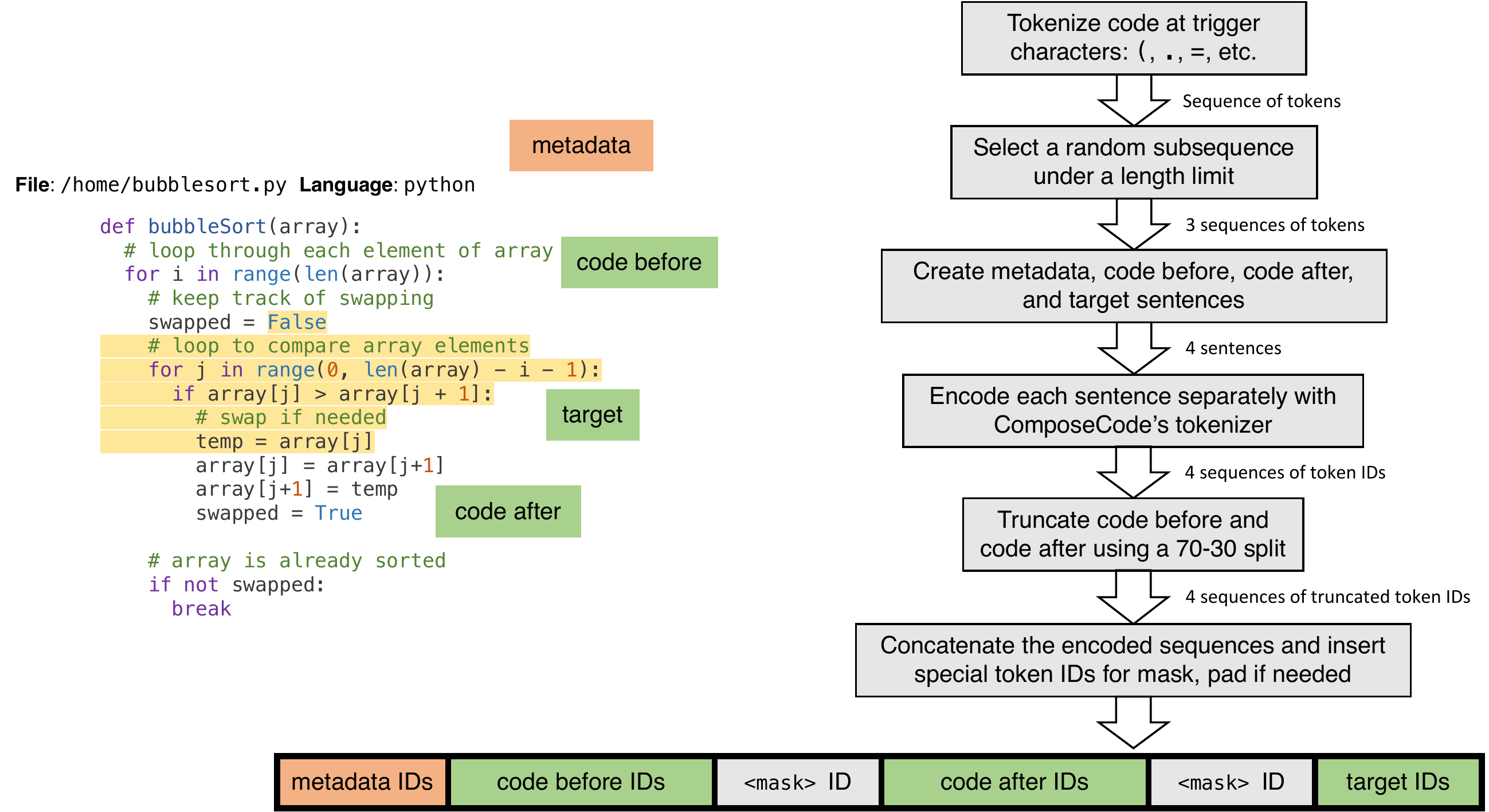}
    \caption{Steps to construct an input to the model in LCM: (i) the code is tokenized at trigger characters where we expect the model to offer suggestions in production, (ii) a random subsequence of tokens is selected to be masked as the ``target'' to predict given the code before and the code after it, (iii) any additional metadata such as the filename is added to the front, (iv) all four strings (metadata, code before, target, code after) are encoded into tokens, (v) since the model's input length is limited, a 70-30 split is applied to the code before and code after if needed, (vi) all tokens are concatenated together into a single list of tokens with special tokens added to denote the masked target portion. The code on the left shows an example with the randomly selected target portion highlighted.}
    \label{fig:lcm}
\end{figure*}

\subsection{Model Architecture and Training Objective}
LLMs largely fall into two main categories.
The BERT~\cite{devlin2019bert} branch of models are trained with the Masked Language Model (MLM) objective that masks out certain parts of the input and trains the model to predict them back.
These models have a notion of bidirectionality, where predicting a token can take into account both preceding and following tokens.
However, MLM is not easily suited for generative tasks.
The GPT~\cite{brown2020language} branch of models are trained with the Causal Language Model (CLM) objective that provides the model with a sequence of tokens and trains it to predict the next token~\cite{brown2020language}.
CLM makes the model more suited for auto-regressive generation tasks, however, it only takes into account the preceding tokens at a given point.

In industrial settings, developers often perform the task of editing where there is code before and after their cursor position.
The code after the cursor contains highly relevant signals about the code being written.
Thus, for a code generation system, we would ideally want both -- a generative model with bi-directionality.
For this purpose, we used the InCoder LLM~\cite{fried2023incoder} as our pre-trained model.
InCoder is a generative model trained with the Causal Masking (CM) objective, where a sequence of tokens are masked out from a given input, appended to the end of the input, and the model is trained to generate the whole sentence left-to-right.
Moreover, the tokenizer used by InCoder tokenizes multiple words into a single token, and as such allows efficient encoding of common code patterns such as \texttt{import numpy as np}.
For more details, we refer the reader to the InCoder~\cite{fried2023incoder} paper.

To suit our use case, we make a few modifications to the CM objective and propose \emph{Language Causal Masking (LCM)}:

\begin{itemize}
    \item CM implements the masking after the text has been tokenized into token IDs, which limits the model during training to only seeing mask spans with edges at common tokenizer tokens. LCM lifts the masking step to the language level and avoids this, similar to the fill-in-the-middle (FIM) task~\cite{bavarian2022efficient}. 
    Also, LCM only masks at certain trigger characters -- that is, characters where the model will be queried during inference such as \texttt{(}, \texttt{.}, \texttt{=}, etc.
    
    \item We prefix certain metadata to the input in LCM, such as the programming language, full path to the file, and the kernel name for notebooks.
    
    \item Through model-level ablations,
    we found an optimal 70-30 split of the model's input length between code before and code after the cursor.
    
    \item Specialized for our use case, LCM has only one mask in any input.
\end{itemize}

A step-by-step overview of constructing an input in LCM is shown in Figure~\ref{fig:lcm}, along with an example code snippet. Once an input is constructed, during training, we maximize the log probability of the language-masked input:
$$
\log \mathcal{P}([\texttt{Metadata}; \texttt{Before}; \texttt{<mask>}; \texttt{After}; \texttt{<mask>}; \texttt{Target}])
$$
where \texttt{Metadata}, \texttt{Before} and \texttt{After} are the tokens in the metadata, code before, and code after the cursor, respectively, \texttt{Target} is the code that was masked, and \texttt{<mask>} is a special token.
During inference, we sample tokens in an auto-regressive manner from the distribution:
$$
\mathcal{P}(\cdot~|~[\texttt{Metadata}; \texttt{Before}; \texttt{<mask>}; \texttt{After}; \texttt{<mask>}])
$$
As we are suggesting lines of code, we stop the generation early once a newline token has been generated.
Due to the real-time nature of our application and the inline suggestion user experience (UX), we only return one sequence of generated tokens.

\subsection{Training data}

InCoder has been trained on (i) a corpus of 159 GB of public code in 28 programming languages with permissive, non-copyleft, open-source licenses from GitHub and GitLab, and (ii) 57 GB of content from StackOverflow questions, answers, and comments.
For training on our internal data, we collected data \company's code repositories and notebooks, i.e., first-party data, applying several filters:

\begin{itemize}
    \item Rather than crawling the entire repository, we used code that is modified through diffs (\company's term for pull requests) checked in by developers as a way of staying close to our end application (i.e., writing code in the IDE). This way we avoid training on code that may have been added a long time ago but is never modified.
    
    \item To keep the training data fresh, we only looked at diffs going back up to 2 years, and only kept the latest versions of files to avoid bugs that may have been patched.
    
    \item For each major target language, we exclude code that is not in production or deprecated. 
    
\end{itemize}

After these filters, our first-party training data included in the order of tens of millions of files amounting to a few billion lines of code across 10+ languages.

We took InCoder-1.3B, the public model with 1.3 billion parameters\footnote{At the time of acceptance of this paper, we have moved to using CodeLlama-7B~\cite{roziere2023code} with 7 billion parameters as our foundation model. See~\cite{Dunay2024FSE-industry} for more details.}, and fine-tuned it further on the above first-party data with our LCM objective. 
For fine-tuning the 1.3B model, we used a batch size of \BatchSizePerDevice per device (\BatchSize effective) and a learning rate of \LearningRate. Training for \NumEpochs epochs with sharded data parallelism took \NumDaysToTrain days on a cluster of \NumTrainingGPUs \GPUType GPUs. We trained the model on all \numLangs languages at once. Following standard practice, the training data is randomized to help the model converge and to ensure that it does not overfit to a single language or dataset. We then deployed the model on a cluster of \NumInferenceGPUs \GPUType GPUs.

\subsection{Model Evaluation Method and Measures}

We describe our methodology to answer RQ1: How well does \cc generate one hidden line of code from existing code snippets?
Our methodology involves a series of evaluation experiments of \cc that try to measure the accuracy of \cc code generation in different languages and the impact of fine-tuning the model with first-party data, i.e., the \company internal codebase. Our evaluation methodology follows a similar strategy to Figure~\ref{fig:lcm}, which we explain in  detail in this section.

The experiments include four steps, namely test dataset collection, random masking, model evaluation, and metric measurements.
Our test set is an unseen random sample of 20K source code files from \company's source code repositories. Unseen samples come from a holdout set of files not used in training. The holdout set is separated by a filepath-based hash to ensure that there is no leakage of data between training and holdout. We collect files that belong to four major programming languages used in \company: Hack, C++, Python, and Flow (JavaScript). Once we collect the test dataset, to replicate code completion scenarios in the IDE, we randomly mask a part of the code snippets ending with the end of a line, with a maximum length of 100 tokens, in each source code file. The length of the masked targets is on average 3\% (or 39 characters) of the context passed to the model. We used heuristics to ensure that the masked part includes nontrivial spans of code, comments, etc. In this scenario, the prediction task is to predict the masked code snippets when the code before and after the masked part are passed to the model as the context. In addition to this context, we pass metadata such as file name, full path, and the programming language. Furthermore, to avoid over inflation of the model accuracy and represent a more realistic use case of \cc code completion, we include extra masking of a random number of tokens after the target line to represent the not yet written part of the developers' code and feed the rest of the code snippet as the context to the model. 
The input context is consistent with the context at the model training, i.e., with the maximum size of 2k tokens and with the 70-30 split between the code before and after the masked target. 
While we use heuristics to make it as realistic as possible, we agree that the backtest may not reflect the actual coding patterns. To address this limitation, we employ multiple evaluation methodologies: backtesting (Section \ref{subec:results-RQ1}), online evaluation on the suggestions made in our production system (Section \ref{sec:resultsRQ2}), and thematic analysis on user feedback (Section \ref{sec:feedback}). 

To evaluate the model, we run different versions of the model including the InCoder model as the public model and the fine-tuned model on the first-party data with the CM objective as well as the fine-tuned one with LCM objective. After running the different versions of the \cc model, we measured two metrics and broke them down into the four main languages of Python, C++, Hack, and JavaScript to evaluate how well the model performs and the impact of fine-tuning with first-party data from \company on the model accuracy. We use the measures of Exact Match (EM), as our most restricted metric, and BLEU score \cite{10.3115/1073083.1073135}, which are used in similar studies \cite{fried2023incoder, lu2021codexglue} and measured by comparing the reference of the masked part of the code and the predicted tokens by the model. The detailed results of this evaluation is described in Section \ref{subec:results-RQ1}.

\section{System Design} \label{design}

\begin{figure*}
    \centering
    \includegraphics[width=0.5\textwidth]{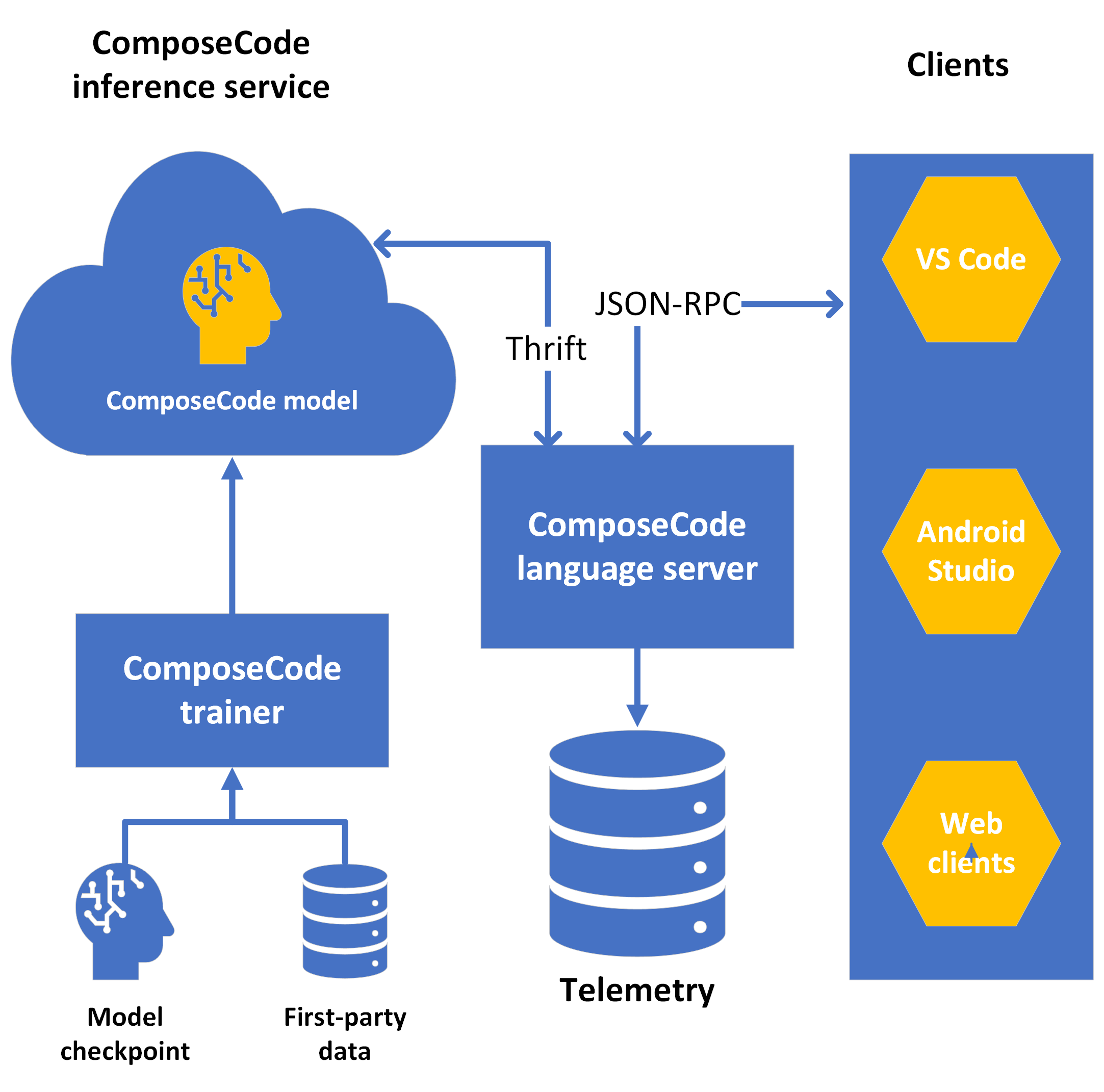}
    \caption{System Architecture for \cc. A pre-trained model checkpoint (InCoder) is trained by the trainer on first-party company data to create the fine-tuned \cc model. This is served in a cluster of inference service machines with GPUs. When clients (VS Code, IDEs, web surfaces) require a code suggestion they send a JSON-RPC request to the \cc language server protocol (LSP), which in turn communicates with the inference service through language-agnostic Thrift calls and sends the generated suggestion back. The LSP also logs telemetry which allows us to compute usage metrics, run experiments, and monitor for regressions. Clients, e.g., VS Code, then display the suggestion through their respective editors.}
    \label{fig:system-architecture}
\end{figure*}

The components of the \cc system are shows in Figure~\ref{fig:system-architecture}. \cc employs a typical client-server architecture in which the \emph{server} is an inference tier that runs the model and the \emph{client} is an editor that surfaces code suggestions, including VS Code and Android Studio. We encode the bulk of the client-side logic in a Language Server Protocol (LSP)~\cite{lsp} conformant language server that is reused across multiple editor integrations.

\subsection{Server}

At \company, we have a tier of machines equipped with \GPUType GPUs, each with sufficient memory to host the fine-tuned \cc model.
Clients can make requests to this tier via Thrift~\cite{slee2007thrift}.
The caller specifies the code before and after the cursor, as well as the file path, language, and metadata to use to process the request.
The request handler is designed to be simple to utilize the most out of the GPUs. It simply processes the request data into a string, tokenizes it, and invokes the model's auto-regressive generation function that runs on the GPU.

Since we are optimizing for latency rather than throughput, each request is processed as soon as it arrives and so there is no inference batching.
Due to latency considerations, when the input length is less than 1500 tokens, we can afford to perform a beam search with a beam width of 3 and return the most likely suggestion.
Otherwise, we perform greedy generation.


\subsection{Language Server Protocol}

To mediate requests between the client and server, we implemented a language server in Rust that we reuse across our various editor integrations. The LSP acts as an orchestrator and handles critical functionalities in the request-response workflow such as calling the inference APIs, displaying the received response as inline completions, logging the telemetry, etc. 

While most LSPs are designed to support a wide array of traditional IDE functionality, such as autocomplete, jump-to-definition, the \cc LSP supports only one meaningful request type: \code{textDocument/inlineCompletions} to help with displaying grey text inline completions.

The LSP also implements core request logic such as debouncing and caching. Debouncing waits for a 20ms pause before re-sending a request to avoid too many requests from being sent as the user is typing in the editor.
Caching saves the response from the server with the context as the key, so that duplicate requests -- which happen frequently as developers type and erase code -- can be avoided.
Moreover, requests that hit the LSP cache are served significantly faster than the service.


\subsection{Clients}

Clients are extensions that run locally on developers' machines and are responsible for ultimately displaying inline suggestions in the editor.
For editors, such as VS Code that support LSP natively, we require relatively little glue code to create the \cc extension. For editors such as Android Studio that does not have native LSP support, we built a small adapter to proxy requests to our LSP.

Furthermore, this architecture makes it straightforward to integrate \cc with our in-house developer tools, such as the web-based variant of Jupyter notebooks.
Because the notebook surface also supports LSP, it was easy to provide our data scientists and ML engineers with the same AI-powered code suggestions as developers working in VS Code.
This ability to plug \cc into any code editing surface internally is an advantage of owning the entire stack.

\subsection{Telemetry}

The LSP is responsible for recording all of the client-side telemetry, which ensures that metrics are recorded consistently across the various surfaces where \cc is used.
Because the LSP receives \code{textDocument/didChange} events, it can calculate most of the metrics of interest without further cooperation from the client.
Nevertheless, the LSP does expect a client to send a custom \code{cc/received} notification upon receipt of the completions so that we can record the full round-trip time from the end user's perspective.
Limiting the demands on the client facilitates integrating \cc into more of our internal tools. 


\subsection{Evaluation Methodology for \cc in Production}
\label{sec:methodProduction}

We used a mixed methods approach~\cite{Creswell2017MixedMethods} to evaluate \cc in production, collecting usage data for RQ2, and feedback comments for thematic analysis for RQ3. 
We follow the principles of Technical Action Research at every phase of the roll out \cite{10.1007/978-3-642-29863-9_17}. This method of doing the rollout steadily in phases helped us measure the effects of \cc in practice  at every step (using the quantitative and qualitative feedback) and iterate on the product experience to improve the means to achieve intended outcomes before rolling out further.
{\bf Our rollout strategy} for \cc consists of gradual deployment in waves of languages: (i) only Python, (ii) Hack, Flow (Javascript), and C++, (iii) others. Within each wave, we rolled it out to increments of 25\% of the developer population until we enable it for 100\% of developers. The rollout was completed after four weeks in Spring of 2023.


{\bf A quantitative methodology} is necessary to answer RQ2: How many suggestions are accepted by engineers and what proportion of the code is written by \cc?
We instrumented telemetry to track various events in the IDE such as displaying a suggestion inline, accepting or rejecting a suggestion, and the length of accepted suggestions. In total, our large-scale deployment resulted in \cc making \numSuggestionsActivity suggestions across \numLangs programming languages. \numDevsActivity distinct developers have seen at least one \cc suggestion. We only count suggestions that were displayed for at least 750 milliseconds to ensure that developers were exposed to a suggestion and had a chance to see and comprehend it.

Our outcome measures are the acceptance rate of suggestions and the percentage of code typed using \cc. These measures have been used in prior work, with, for example, Google reporting that 3\% of the code typed by engineers was from their AI~\cite{googleblog}.

{\bf A thematic analysis} is necessary to answer RQ3: How do developers perceive \cc in their daily work?

We collected data by manually coding user posts in the \cc feedback group. The first two authors coded responses and discussed them until a consensus was reached. In total, 70 engineers provided feedback. As starting codes, we used the standard positive, CSAT (Customer Satisfaction), or negative, DSAT (Customer Dissatisfaction). We then broke positive and negative feedback into subcodes. We then grouped codes and those that achieved saturation were abstracted into themes. For the major themes, we report the frequency of the themes. To ground our themes in the evidence, we provide representative quotations of both positive and negative feedback in Sections~\ref{sec:PositiveExperiences} and \ref{sec:NegativeExperiences}, respectively. This feedback also allowed us to reflect on \cc and to identify the major factors affecting its perceived usefulness.

\section{Results}
\label{eval}

\subsection{RQ1. Model Evaluation Results}
\label{subec:results-RQ1}
\textit{How well does \cc generate one hidden line of code from existing code snippets?}

Before we can release \cc into production, we need to ensure that the model is sufficiently accurate. 
We simulate an AI-authoring experience by selecting a random line from a file and providing the surrounding context to \cc to determine how well it would have generated the hidden line. 
The training of the model and description of the metrics and evaluation methodology are in Section~\ref{sec:methodModel}. We report the results for the number of exactly matching lines as well as the BLEU score for hidden lines in Table~\ref{tab:EM-BLEU}. 

\begin{table*}[h]
\centering
\caption{Outcome metrics across programming languages for the Public model and the fine-tuned models. Our LCM fine-tuned model substantially outperforms the Public model by between 1.4 times and 4.1 times.}
  \begin{tabular}{lrr|rr|rr|rr}

\toprule
  \multicolumn{9}{c}{Percentage of Exact Matches (improvement over Public Model by $\times$ times)}	\\	\hline															
Models	&	\multicolumn{2}{c}{Python}			&	\multicolumn{2}{c}{Hack}			&	\multicolumn{2}{c}{Flow}			&	\multicolumn{2}{c}{C++}			\\ \hline
Public	&	23.1\% &	$-$	&	14.2\% &	$-$	&	18.6\% &	$-$	&	23.4\% &	$-$	\\ 
Fine-tuned (CM)	&	35.5\% &	1.5$\times$ &	37.8\% &	2.7$\times$ &	30.8\% &	1.7$\times$ &	24.2\% &	1.0$\times$ \\ 
Fine-tuned (LCM)	&	48.8\% &	2.1$\times$ &	57.7\% &	4.1$\times$ &	52.2\% &	2.8$\times$ &	40.0\% &	1.7$\times$ \\ \hline \hline
\multicolumn{9}{c}{BLEU Score (improvement over Public Model by $\times$ times)}	\\ \hline																
Models	&	\multicolumn{2}{c}{Python}			&	\multicolumn{2}{c}{Hack}			&	\multicolumn{2}{c}{Flow}			&	\multicolumn{2}{c}{C++}			\\ \hline
Public	&	37.9\% &	$-$	&	27.6\% &	$-$	&	33.8\% &	$-$	&	39.9\% &	$-$	\\ 
Fine-tuned (CM)	&	50.0\% &	1.3$\times$ &	52.6\% &	1.9$\times$ &	45.8\% &	1.4$\times$ &	40.3\% &	1.0$\times$ \\ 
Fine-tuned (LCM)	&	62.0\% &	1.6$\times$ &	73.0\% &	2.6$\times$ &	68.6\% &	2.0$\times$ &	56.2\% &	1.4$\times$ \\

    \bottomrule
  \end{tabular}
  \label{tab:EM-BLEU}
\end{table*}

We compared the performance of various models such as the public checkpoint of the InCoder model and the fine tuned versions with different objectives. The Public model performs reasonably well. In Table~\ref{tab:EM-BLEU}, we see that the percentage of exact matches ranges from 14\% for Hack to 23\% for C++. The BLEU score is also reasonable ranging from 28\% for Hack to 40\% for C++. 

Table \ref{tab:EM-BLEU} also shows the importance of fine-tuning the \cc model on \company source code so that it can learn \company-specific code styles, formatting, and libraries. The LCM model always outperforms the CM model. The jump is dramatic especially in the case of LCM, with Hack seeing a 4.1 times improvement on the Exact Match and a 2.6 times improvement on the BLEU score over the Public model. The increase is still substantial for C++ with corresponding improvements of $1.7\times$ and $1.4\times$, respectively. 

This experiment provides strong evidence that an internally fine tuned model outperforms an off-the-shelf model that is trained on external data only. The LCM model is very successful and is used in the subsequent sections where we roll out \cc to developers.

\begin{tcolorbox}

We find that LCM \cc model exactly recreates the masked line between 40\% and 58\% of the time and has a high BLEU score between 56\% and 73\%, depending on the programming language. This represents an improvement over the Public model between $1.4\times$ and $4.1\times$. 
 
\end{tcolorbox}

\subsection{RQ2. Adoption Results}
\textit{How many suggestions are accepted by engineers and what proportion of the code is written by \cc?}
\label{observations}
\label{sec:resultsRQ2}

We discuss the design of our \cc system and its integration with VS Code and other editors used at \company in Section~\ref{design}. All changes at \company must be gradually rolled out to users and we discussed our rollout strategy and adoption metrics in Section~\ref{sec:methodProduction}. 
Table~\ref{tab:NumSuggestions} shows the adoption rates of \cc. 

\begin{table*}[h]
\caption{The acceptance rate and the proportion of code that was written by \cc across programming languages}
\centering
\begin{tabular}{lrrrr}
\toprule 
Language & \# Suggestions shown & Acceptance rate & Code by \cc & \# Users \\
\midrule
Python & 1.9M & 22\% & 8\% & 10.7K \\
Hack & 1.3M & 22\% & 10\% & 5.5K \\
C++ & 608.1K & 20\% & 10\% & 2.5K \\
Flow (Javascript) \cite{FlowJs} & 583.2K & 18\% & 7\% & 2.5K \\
Rust & 74.2K & 17\% & 9\% & 212 \\
Objective C++ & 57K & 18\% & 6\% & 429 \\
Objective C & 34K & 18\% & 6\% & 299 \\
C & 23.5K & 21\% & 12\% & 201 \\
Typescript & 8.9K & 19\% & 10\% & 76 \\
\bottomrule
Total & 4.5M & 22\% & 8\% & 16K \\ \hline
\end{tabular}
\label{tab:NumSuggestions}
\end{table*}

\cc made \numSuggestionsActivity suggestions across \numLangs programming languages. \numDevsActivity distinct developers have seen at least one \cc suggestion. As shown in Table \ref{tab:NumSuggestions}, we observe a suggestion level acceptance rate of \accRateActivity across \numLangs programming languages for suggestions that were displayed for at least 750 milliseconds. Imposing a lower bound on the display time of the suggestions when calculating the acceptance rate helps in making sure developers who were exposed to a suggestion had a chance to see and comprehend it \cite{googleblog}. 

Our rollout strategy started with Python, which explains why there are more recommendations in that language. However, even typescript has over 8.9K recommendations and an acceptance rate of 19\%, which is a substantial sample from a statistical perspective. \cc has been rolled out to 100\% of engineers at \company.

We also log the length of each accepted suggestion, and the number of characters developers type when authoring code (excluding large copy-pastes, refactors, etc.).
This allows us to calculate, at a character level, the percentage of the code typed by developers that came through accepting \cc suggestions, which we computed to be \percCodeActivity. \cc is writing a substantial portion of the codebase, and in the next section we use developer feedback to understand what types of code are being accepted.





\begin{tcolorbox}

We make over 4.5M suggestions to 16K engineers. We see an acceptance rate of 22\% which is comparable to those at Google and GitHub. We see that 8\% of the changed code is written by \cc, which is higher than the 3\% reported by Google. \cc is rolled out to 100\% of engineers at \company.
    
\end{tcolorbox}

\subsection{RQ3. Developer Feedback Results}
\label{sec:feedback}
\textit{How do developers perceive \cc in their daily work?}

At \company each internal tool has a feedback forum. The first two authors conducted a thematic analysis of the responses from 70 engineers (the methodology is described in Section~\ref{sec:methodProduction}). Similarly, the model does not differentiate between suggestion types such as code suggestions and documentation suggestion when making suggestions. We report the frequency of themes and suggestion types in Table \ref{Tab:Quantative-Feedback-Responses}. In subsequent sections, we provide grounded evidence of the themes through quotations and discuss both positive and negative feedback.

\begin{table*}[h]
\caption{Frequency of themes coded from the feedback of 70 engineers. There are only six unfavorable responses, so we discuss each in Section~\ref{sec:NegativeExperiences}}
\begin{tabular}{*{2}{p{5cm}} l} 
 \toprule
 Theme & Description & Engineers \\ \hline
 Accelerate coding & \cc helped speed up my coding process & 14 (20.0\%) \\
Discovery & \cc helped me discover APIs, write boilerplate code faster & 16 (22.8\%)  \\
Documentation & \cc helped me with generating in-code documentation, docstrings, etc. & \hspace{.1cm} 5 (\hspace{.1cm}7.1\%) \\
Suggestion accuracy & \cc suggestions are accurate and not noisy & 11 (15.7\%) \\
Generally favorable & \cc is generally great, and I will keep using it for my coding tasks & 18 (25.7\%) \\ \hline
Unfavorable & \cc is not useful & \hspace{.1cm} 6 (\hspace{.1cm}8.5\%) \\ \hline \hline
Total & & \hspace{0.04cm}70 (100\%) \\ 
\bottomrule
\end{tabular}
\label{Tab:Quantative-Feedback-Responses}
\end{table*}

Table \ref{Tab:Quantative-Feedback-Responses} lists the distribution of qualitative feedback responses. 91.5\% of the \cc users gave a favorable response while 8.5\% of them gave an unfavorable response. Many respondents (15.7\%) appreciated the fact that \cc suggestions are accurate and \cc was able to auto-complete the code they were going to type. 

In addition to the accuracy of suggestions, 23\% of the users said \cc helped them discover new APIs, automating tedious tasks such as boilerplate code. Therefore, \cc is not just helping with typing code faster (as demonstrated in Table \ref{tab:NumSuggestions}) but also saving time and effort people spend searching for APIs and documentation.

\cc is helping people move faster with with 20\% of the users stating that they found \cc to accelerating their coding activity. In addition to helping write code faster, \cc is helping developers produce more in-code documentation and API documentation. This is an interesting side-effect and demonstrates the model's ability to generate text as an output from code in addition to code generation.

\subsubsection{Evidence of Positive Experiences}
\label{sec:Favorability}
\label{sec:PositiveExperiences}

After analyzing all the qualitative feedback, we identified the following
main factors that make a developer inclined towards finding a system like \cc useful.

\textit{Favorable scenarios for \cc:} The scenarios for which \cc was able to add the biggest value includes (but is not limited to), auto-completing lines of code, API discovery, boilerplate coding, suggesting standard libraries, generating in-code documentation, etc. 

When it comes to developers, the developers who benefited the most are the ones who work on authoring code that involves writing boilerplate code and idioms, the ones who follow the typical coding patterns employed at \company, and the ones that employ standard first-party (or third-party) libraries to accomplish their coding tasks. Additionally, \cc is well received by the developers who tend to work on building pipelines and common infrastructure tasks that involve writing heavily templatized code. This can be corroborated by the distribution of responses listed in Table \ref{Tab:Quantative-Feedback-Responses} and the representative quotes listed below.

\textbf{Grounded evidence of positive experiences.} 
We share quotations that exemplify the positive experiences of developers. A developer who got access to \cc shared their first impressions:
\begin{quote}
\textit{"Today I needed to write a quick tool for {\normalfont [task]}. I opened VS Code and started writing {\normalfont [code]}. Suddenly this thing called \cc popped up and wrote all the code for me! This was a super delightful experience, it really made my day, so thank you!"}
\end{quote}

In this case, the developer was unaware of the existence of \cc. This quote summarizes the delightful coding experience that the developers at \company experienced when they pair-program with \cc.

Several remarks are clear indicators of the usefulness of employing \cc in code authoring workflows.
\begin{quote}
    \textit{"I have been a big fan of {\normalfont [\cc]} ever since I got access. I do not think I have been in any coding session since then where I did not use it."} \\
    
    \textit{"The suggestions {\normalfont [from \cc]} are quite relevant and do not show up too often -- not necessarily ``rarely'', but not obnoxiously often either. It is a nice balance."}\\
    
    \textit{"I was blown away that after writing the following {\normalfont [code]}, \cc successfully predicted the correct completion. This is amazing because I could not remember the name of the global variable or where it was in the code, but \cc knew what it was."}
\end{quote}

Many developers highlighted the fact that the predictions are relevant. Also, we received feedback about how nicely \cc navigates the precision versus recall problem by not showing suggestions too often.

Several pieces of feedback highlighted \cc's ability to help developers discover new APIs or ramp them up quickly on unfamiliar APIs.
\begin{quote}
    \textit{"I wanted to post about a great experience I had in using \cc in the {\normalfont [internal]} codebase. I have been getting back into coding and was rusty on the different arguments that exist in the {\normalfont [library]} operators. Some operators can have up to {\normalfont [hundreds of]} arguments so this is a challenging problem. \cc has helped me several times in a day by giving me the right argument that I cared about, whereas my previous workflow would have been to {\normalfont [search]} for the operator and spend several minutes on each."}\\
    
    \textit{"\cc has been a game changer for working with APIs that I have not used before. Even if it is not exactly what I need, it often shows me approximately what I should pass or how I should call it. This is much faster than looking for other call sites to figure out how to use this {\normalfont [API]}."}
\end{quote}

We also observed interesting side effects such as increased in-code documentation when developers use \cc to make code changes. As \cc provides accurate descriptions of code changes and a template to start with, developers tend to accept the suggestions, make changes (as necessary), and push the documentation changes (along with the source code changes) in their diffs. This is reflected in some of the anecdotes listed below.

\begin{quote}
    \textit{"I find \cc particularly useful when writing docstrings. Without \cc, I would not even imagine that I can write {\normalfont[so many]} lines of docstring for my actual code."}\\
    
    \textit{"I really like how \cc highlights the value of naming and the quality of documentation, which used to be mainly for long-term benefit, but now good naming and {\normalfont [documentation]} gives you an immediate effectiveness boost."}
\end{quote}

\subsubsection{Evidence of Negative Experiences} 
\label{sec:NegativeExperiences}
After analyzing the qualitative feedback, we found that \cc is not helpful in scenarios where developers use specialized APIs and libraries. \cc seems to be struggling with suggesting the correct code when the developers employ atypical coding patterns. 

Some developers found the traditional, semantic, auto-complete functionality to be more useful in those cases. Developers also complained about the possibility of \cc to hallucinate \cite{guerreiro2023hallucinations, li2023dark} when recommending the uncommon API names, URLs, etc. 

Developers also found the coexistence of \cc with traditional auto-complete to be vital. Sometimes, when these two systems compete to show suggestions, it creates a disruptive experience for developers. Also, commonly used keyboard shortcuts such as ``Tab'' are overloaded to accept \cc suggestions and traditional auto-complete suggestions. Solving these UX problems is vital to facilitating a smoother integration with existing semantic engines, which is addressed in our discussion Section \ref{subsec:ux}. 

\textbf{Grounded evidence of negative experiences.} A few developers disabled \cc as they found \cc to be noisy and intrusive. We list a few anecdotes below. 

\begin{quote}
    \textit{"\cc is good, however, it seems to reduce my coding speed. The reason is that the traditional auto-complete was typically suggesting {\normalfont [real]} functions, but \cc is suggesting hallucinations. These are not always correct and I would like a UX design where I could switch between traditional auto-complete and \cc."}\\
    
    \textit{"I have realized that {\normalfont [\cc]} tends to struggle while working with function signatures. The suggestions it offers do not match the actual signature, although they make sense."}\\
    
    \textit{"The {\normalfont [code editing]} experience is sometimes good and sometimes bad. Auto-complete struggles to work together with \cc~-- they seem to be competing with each other."}
\end{quote}

This highlights the importance of investing in UX research to deliver the best experience for AI-assisted code authoring while coexisting with traditional auto-complete systems. The feedback also talks about the problems faced by LLMs concerning hallucinations and grounding \cite{10.1145/3571730, guerreiro2023hallucinations}. We are actively exploring this area to make \cc a productive experience for all the developers.

\begin{tcolorbox}

The feedback was overwhelmingly positive, with only 8.5\% of comments containing negative feedback and less than 1\% of engineers turning off \cc. The top two feedback comments related to discovering API and dealing with boilerplate (23\%), and accelerating coding (20\%). \company continues to invest in \cc efforts to iteratively improve the AI-assisted authoring experience.

\end{tcolorbox}

\section{Threats to validity} \label{threats}


\subsection{Generalizability}

\cc is an entirely internal tool that has only been deployed at \company.
Therefore, there is a threat of drawing any general conclusions from the quantitative metrics or qualitative feedback presented.
It is possible that the results might not hold elsewhere, and for this reason, we cannot assume a priori that other AI-based coding assistants built outside \company will perform similarly.
Similarly, the value-add of fine-tuning the LLM on \company's internal code (Table~\ref{tab:EM-BLEU}) may also not translate externally.
However, we see evidence of the need to fine-tune public models on internal data at Google~\cite{googleblog}. Without this  tuning on internal APIs and functions, the model would be unable to make reasonable suggestions for the internal codebase. 

\subsection{Construct Validity}

There are many AI model evaluation metrics. We have selected the commonly used BLEU score and the percent of exact matches. While there are limitations with these metrics, exact match is a very conservative measure only succeeding when the code is perfectly generated.

There have been few deployments of AI reported in the literature, so we used standard measures employed at \company, such as the level of adoption and the acceptance rate to quantify the efficacy of AI-authoring at scale. We triangulated these numerical results with a thematic analysis of feedback from 70 developers to ensure that the model's success resulted in a usable product. 

\subsection{Internal Validity}
While we have taken significant measures to reduce bias in the results presented (e.g., by waiting several weeks after randomized rollout), we cannot guarantee that there is absolutely no bias.
The time period chosen for measuring the metrics in Table~\ref{tab:NumSuggestions} could have had external events beyond our control that influenced developers' coding behaviors during that time.
Some languages such as Typescript only has a small number of developers, who might have had an outsized influence on the metrics for those languages.

Qualitative feedback, in general, comes with a degree of subjectivity.
While \company encourages a culture of open feedback, developers might not have been comfortable sharing negative assessments with colleagues.
We facilitated easy opt-out, but less than 1\% had opted out, indicating that a majority of developers found \cc to be a net positive experience.

The categorization of the qualitative feedback into different categories was done manually using a basic thematic analysis rather than a full grounded theory study. We tried to reduce subjectivity in this process by making two authors in the team categorize them and then reconcile the different labels.
Nevertheless, a different set of people or more systematic methodology might have produced a different categorization in Table~\ref{Tab:Quantative-Feedback-Responses}.

\section{Related Work} \label{related}
In this section, we review the related work and position our contribution in the literature. 
Neural program synthesis is a rapidly evolving area with a large body of work discussing the possibility of applying neural architectures, such as Bayesian networks, and Transformers, to solve program synthesis tasks \cite{bruch2009learning,robles2008program,proksch2015intelligent,zhou22improving, kim21code, chen2021evaluating}. These models are typically evaluated against public data sets such as MBPP (974 Python programming tasks, designed to be solvable by entry-level programmers) \cite{austin2021program} and Human Eval (164 original human-curated Python programming problems) \cite{chen2021evaluating}. The data set we used to evaluate the accuracy of code generation contains 20,000 data points sourced from real code bases, curated across \numLangs programming languages, at \company.

There has been a limited deployment of AI-assisted code generation systems in large-scale industrial environments with tens of programming languages and thousands of developers using them daily \cite{intellicode, codewhisperer, copilot, googleblog}. Availability of research in the public domain is even limited when it comes to evaluating the accuracy of neural program synthesis techniques on large scale industrial data sets and understanding the effects of such tools on software developers using mixed methods studies, 


The closest in spirit to our work is the brief blog post from Google \cite{googleblog}. At Google, deploying a hybrid semantic ML code completion to 10k+ Google employees compared to a control group observed a 6\% reduction in coding iteration time (time between builds and tests) when exposed to single-line ML completion. They reported that 3\% of new code (measured in characters) was generated from accepting ML completion suggestions. The acceptance rate for single-line code completions is 25\%. These results demonstrate the impact on developer productivity. In the OSS domain, GitHub CoPilot reported an acceptance rate of 30\% for code completions \cite{GHCP-Acc}.

Our more detailed findings are in line with those found by Google and GitHub with a 22\% acceptance rate for \numSuggestionsActivity single-line code completions across \numLangs programming languages. At the same time, our suggestions accounted for 8\% of the code written by engineers at \company compared to the 3\% reported by Google.

There have been some empirical evaluations of GitHub’s Copilot \cite{copilot} in actual use for automatic code completion. In a grounded theory study of 20 developers using Copilot, Barke \etal \cite{Barke2023OOPSLA1} found that developers switched between coding acceleration and discovery use cases. Our engineers also reported these same themes as their most common positive feedback. Nguyen et al. \cite{nguyen2022empirical} used 33 LeetCode questions to create queries for Copilot in four different programming languages. They found that Copilot's Java suggestions have the highest correctness score (57\%) while JavaScript is the lowest (27\%) \cite{nguyen2022empirical}. While LeetCode examples are useful for algorithmic interview preparation, they differ dramatically from normal engineering work.

Code generation could also be achieved by prompting a general LLM such as ChatGPT~\cite{liu2023improving}. Also, it has been found that different types of prompts can lead to different code generation performance. We performed prompt engineering on the fine-tuned model (for e.g., determining the 70-30 split of the context fed to the model) to make it effective for code suggestions. We also show in Table~\ref{tab:EM-BLEU} that public models underperform on \company’s internal code base for code suggestion tasks compared to models fine-tuned on it.

A study with 15,000 synthetic data points curated from a C\# code base from an industrial software development environment reported an accuracy of 30.7\% for predicting single tokens, such as identifiers, when employing a modified Best Matching Neighbor approach (BMN\#) \cite{8812116}. Offline evaluations of Neural methods for auto-completion in a comparable industrial setting yielded accuracy of 41.9\% on languages with heavily customized frameworks such as Hack \cite{zhou22improving} when tested on a couple of thousands of examples and when tested on open-source data sets (CodeSearchNet) \cite{husain2020codesearchnet} yielded an Exact Match score of 20.10\% and BLEU score of 32.79\% \cite{lu2021codexglue}. Our study of 20,000 data points curated, across \numLangs programming languages, from \company's production code base shows that \cc achieves the highest correctness score of 52\% and BLEU score of 73\% for Hack language and lowest correctness score of 40\% and BLEU score of 56.2\% for C++.

While Copilot is impressive, not all usage of has been positive. A qualitative study  with 24 participants to understand how programmers use and perceive Copilot found that it did not necessarily improve the task completion time or success rate by a significant margin \cite{vaithilingam2022expectation}. However, 19 of 24 participants (79\%) answered that they preferred Copilot over Intellisense. A similar qualitative study of \cc with a larger group of software engineers at \company (70 participants) resulted in a 91.5\% favorable feedback. Developers cited reasons such as accelerated coding or typing, helping in discovering new APIs and writing boilerplate code in unknown programming languages, helping with generating documentation for their code, accuracy of suggestions and its ability to predict what the developers are going to type as the main reasons.

\section{Discussion and Lessons Learned}
\label{sec:challenges}
\label{sec:discussion}

We organize our discussion around the challenges and learnings in creating trust, making the user experience better, and employing metrics to evaluate AI coding assistants at scale.

\subsection{Trust}
\label{subec:trust}

Most of today's code generation LLMs consider code as a sequence of tokens, similar to natural language.
That is one of the reasons for the observed acceleration in the model-building process -- dealing with the complexities of parsing source code or developing semantic understanding is not required anymore.
However, this has some side effects: most of the models cannot guarantee that the generated code compiles, is syntactically correct, or executes.
For example, the model may make a ``suggestion'' by generating an API call that does not exist.
Making developers trust and accept suggestions in this setting is a huge challenge.
Other factors can also impact trust such as the source corpus used for training and its validity, biases in training data, security issues, vulnerabilities, constant data drift, etc.


The traditional precision versus recall problem in machine learning also plays into trust.
If the model is optimized for recall and generates a lot of incorrect suggestions, developers will lose faith in the system as it becomes noisy.
If the model is optimized for precision and generates only a few accurate suggestions, developers might eventually stop caring about the system as it becomes sparse.
Also, it will be hard to justify the return on investment if the system is not generating enough suggestions.
Studies\cite{githubblog} show that developers are fine with reworking a suggestion as long as the model provides a useful starting point or structure with relevant libraries and function calls.
However, requiring a lot of rework might contribute to the reduction of trust and confidence in the model.

\paragraph{Learnings for \cc.}
Building trust was an important part of our productization, as it is directly affected by model accuracy.
First, we worked with language partner teams at \company to identify obsolete patterns (e.g., ``PHP-isms'' in Hack) and not-in-production code (e.g., experiments) in the codebase and filter it out from our training data.
In a similar vein, we only train on code that is being actively modified, that is, containing commits in the last two years, as opposed to all code in the repository.

Secondly, from developer feedback, we found out that contextual information adds significant value to the suggestion accuracy, such as the code \emph{after} the cursor, the file being edited, or the kernel being used (in a notebook).
We modified our model's training objective to take into account this contextual information (for more details see Section~\ref{model}).

Finally, our rollout strategy allowed us to incrementally build trust in the product (details about our rollout strategy is explained in Section \ref{sec:methodProduction}).

\subsection{User Experience}
\label{subsec:ux}

Due to the generality of LLMs, code generation can be done at multiple levels of granularity: token completion, statement completion, completing multiple lines, or generating entire blocks.
Depending on factors such as suggestion confidence, user context, and task context, we will need to generate suggestions at different granularities.
For example, when developers are in their coding ``flow'', suggesting a large block of code and moving their existing code down -- and moving it back up when they type over the suggestion -- results in an extremely jarring experience.
This is because developers would constantly need to switch between typing code and {\it reviewing} the suggested block of code, resulting in an unpleasant ``code review on the fly'' situation.
Therefore, when to suggest, what to suggest, and how much to suggest is an essential part of the user experience.



Similarly, performance and latency are also an important consideration.
Coding assistants operate in a real-time environment with strict requirements on latency.
Developers do not want to wait for several seconds for the system to generate a single suggestion as they are typing it -- the system has to match the developer’s typing speed.
Studies\cite{hofmann14eyetracking} have shown that sometimes users adjust their typing behaviors (and speed) to match autocomplete suggestions in IDEs.
However, slowing oneself down should be compensated by passing the right suggestions and making it a ``net positive'' experience.

\paragraph{Learnings for \cc.}
After several iterations, we found, through developer feedback, that offering suggestions one line at a time -- that is, completing the current line of code where the cursor is -- strikes a good balance between suggesting an adequate amount of code and avoiding the jarring effect.
Furthermore, even within the current line, we do not show suggestions if there is any code to the right of the cursor, except certain tokens such as ), \}, and ].

In terms of latency, developers were fine with suggestions appearing within 300ms - 500ms, and certainly not beyond 1s. We observed that the acceptance rate went down when the end-to-end latency went up. 
To bring down the end-to-end latency below this threshold, we employed a suite of optimizations such as caching and debouncing (for more details see Section~\ref{design}).
With this experience, developers who are looking for {\it multi-line suggestions} simply press Tab as many times as needed.
This allows them to edit the currently suggested line before going to the next line, which makes the subsequent suggestions more accurate.
They can also press a manual keyboard shortcut to request a multi-line suggestion, but we found that developers rarely use this workflow.

\subsection{Metrics}
\label{subsec:metrics}

Evaluating the usefulness of AI-generated suggestions is a major challenge\cite{10.1145/3520312.3534864}.
This includes, but is not limited to, the complexities involved with the granularity of suggestions, developers' flexibility to rework the generated code, defining true positives, etc.
A strict metric like Exact Match, BLEU score or percentage of accepted suggestions as-is will underrepresent the actual benefits offered by these solutions.
For example, while the acceptance rate is a good metric to measure product usage, a high acceptance rate does not necessarily mean developers are being more productive when using the tool -- that has to be measured explicitly.

Furthermore, evaluation needs to go beyond just measuring product impact -- it should help us understand the cases where the suggestions are inaccurate, or what the pain points are with respect to the user experience, to address those issues.
Additional complexities including privacy, compliance, and proprietary data make it harder to track all the telemetry that is required to evaluate or monitor the product.


\paragraph{Learnings for \cc.}
We developed a suite of metrics that can quantify our impact, some of which are standard definitions of internal metrics to measure the scale of deployment of tools at \company.
We track usage metrics such as the number of active users, user engagement and retention, number of suggestions offered, acceptance rate, and the percentage of code typed by developers, that comes from accepting \cc's suggestions (for more details see Section~\ref{eval}).

\section{Conclusion} \label{conclusion}

In this paper, we introduced an AI-based coding assistant system named \cc, discussed how we scaled it to \numDevsActivity developers across \numLangs programming languages at \company, and presented metrics and feedback to understand the impact of the system on code authoring. 

We made the following contributions:

We then presented the system architecture of \cc. We built \cc using three primary components: the model inference service, Language Server Protocol (LSP), and the client. When a developer types code, the service receives a request to perform inference and generates a sequence of tokens to auto-complete the statement(s). The LSP and the client help orchestrate the inference calls and implement the logic to display the suggestion.

We presented details about the underlying InCoder-based LLM that powers \cc.
We introduced a custom training objective, Language Causal Masking, that suits our application of suggesting individual lines of code.
In doing so, we conducted an offline evaluation that showed a $1.4\times$ and $4.1\times$ improvement brought about by fine-tuning the LLM on the \company's internal codebase.

We presented quantitative metrics that show the scale and impact of \cc.
\numSuggestionsActivity suggestions were shown to developers resulting in \percCodeActivity of code typed by users of \cc coming from its suggestions.

We presented qualitative feedback from developers.
We found that an overwhelming 91.5\% of the feedback shared by developers was favorable towards \cc, indicating that it adds significant value to their coding experience.
Through a thematic analysis of the feedback, we found, for example, that \cc helps developers discover unfamiliar APIs and write documentation. \cc is now rolled out to all developers, and less than 1\% of developers disabled it.  

In the future, we plan to enable more features such as
building dynamic context from other files to improve suggestion accuracy, incorporate feedback from user actions on suggestions to perform inference-time optimizations, more interaction modalities such as a conversational modality within the IDE, functionality to explain source code and provide code walk-throughs, etc. We are exploring opportunities to leverage semantic information to perform pre-processing and post-processing to improve suggestion accuracy by reducing hallucinations. Furthermore, we see opportunities to expand this technology, beyond code authoring, to help developers across the software development life cycle.

\bibliographystyle{ACM-Reference-Format} 
\bibliography{Paper}

\end{document}